\documentclass[twocolumn,apl,epsf,superscriptaddress]{revtex4}
\usepackage{amsmath}
\usepackage{amsfonts}
\usepackage{epsfig}
\usepackage{epsf}
\usepackage{array}
\usepackage{color}
\usepackage{ulem}

\newcommand{\vect}[1]{\mathbf #1}
\newcommand{\vecs}[1]{\mbox{\boldmath $#1$}}

\begin{document}

\title{Critical Spin Fluctuation Mechanism for the Spin Hall Effect}

\author{Satoshi Okamoto}
\altaffiliation{okapon@ornl.gov}
\affiliation{Materials Science and Technology Division, Oak Ridge National Laboratory, Oak Ridge, Tennessee 37831, USA}
\author{Takeshi Egami}
\affiliation{Materials Science and Technology Division, Oak Ridge National Laboratory, Oak Ridge, Tennessee 37831, USA}
\affiliation{Department of Materials Science and Engineering, The University of Tennessee, Knoxville, Tennessee 37996, USA}
\affiliation{Department of Physics and Astronomy, The University of Tennessee, Knoxville, Tennessee 37996, USA}
\author{Naoto Nagaosa}
\affiliation{Department of Applied Physics, The University of Tokyo, Bunkyo-ku, Tokyo 113-8656, Japan}
\affiliation{RIKEN Center for Emergent Matter Science (CEMS), Wako, Saitama 351-0198, Japan}

\begin{abstract}
We propose mechanisms for the spin Hall effect in metallic systems arising from the coupling between conduction electrons and local magnetic moments
that are dynamically fluctuating.  
Both a side-jump-type mechanism and a skew-scattering-type mechanism are considered. 
In either case, dynamical spin fluctuation gives rise to a nontrivial temperature dependence in the spin Hall conductivity. 
This leads to the enhancement in the spin Hall conductivity at nonzero temperatures near the ferromagnetic instability.  
The proposed mechanisms could be observed in $4d$ or $5d$ metallic compounds. 
\end{abstract}


\maketitle

\date{\today }


{\it Introduction}.---%
The spin Hall (SH) effect is the generation of spin current along the transverse direction by an applied electric field \cite{Dyakonov1971,Hirsch1999}. 
Because it allows us to manipulate magnetic quanta, i.e., spins, without applying a magnetic field, 
this would become a key component in creating efficient spintronic devices. 
By combining the SH effect and its reciprocal effect, the inverse SH effect \cite{Saitoh2006}, 
a variety of phenomena have been demonstrated (for recent review, see Refs.~\cite{Murakami2011,Sinova2015}). 
As in the anomalous Hall effect ~\cite{Nagaosa2010}, 
the relativistic spin-orbit coupling (SOC) plays the fundamental role for the SH effect,  
and both intrinsic mechanisms \cite{Sinova2004,Murakami2004} and extrinsic mechanisms \cite{Smit1955,Berger1970,Crepieux2001,Tse2006}
have been proposed. 
Whereas many theoretical studies considered static disorder or impurities at zero temperature, 
the effect of nonzero temperature $T$ in the SH effect has been addressed using phenomenological electron-phonon coupling \cite{Gorini2015,Xiao2018} 
or first-principle scattering approach \cite{Wang2016}.

At present, the intensity of the SH effect is too weak for practical applications \cite{Hoffmann2013}. 
One of the pathways to enhance the spin-charge conversion efficiency or the SH angle $\Theta_{SH} = \sigma_{SH}/\sigma_c$,
where $\sigma_{SH(c)}$ corresponds to the SH (charge) conductivity, is to reduce the charge conductivity $\sigma_c$. 
For example, Ref.~\cite{Fujiwara2013} proposed to use $5d$ transition-metal oxides, IrO$_2$, where the strong SOC comes from Ir, rather than metallic materials. 
The SH effect in the surface state of topological insulators with spin-momentum locking has been also studied \cite{Ong2018}.  
More recently, Jiao {\it et al.} reported the significant enhancement in SH effect in metallic glasses at finite temperatures~\cite{Jiao2018}. 
Because such enhancement is not expected in crystalline systems \cite{Vila2007}, 
it was suggested that local structural fluctuations \cite{Gorini2015,Karnad2018} are responsible for this effect, similar to the phonon skew-scattering mechanism. 
Thus, the fluctuations of lattice or some other degrees of freedom at finite temperatures 
could provide a route to improve the efficiency of the SH effect. 

For magnetic systems, the effect of finite temperatures has been studied for the anomalous Hall effect in terms of skew scattering \cite{Kondo1962} and 
resonant skew scattering \cite{Fert1972,Coleman1985,Fert1987}. 
Theories for the resonant skew scattering were further developed by considering strong quantum spin fluctuations 
for systems with the time-reversal symmetry (TRS), therefore for the SH effect rather than the anomalous Hall effect  \cite{Guo2009,Gu2010a,Gu2010b}. 
Later, the relation between the anomalous Hall effect below the ferromagnetic transition temperature $T_C$ and the SH effect above $T_C$ was investigated 
by including nonlocal magnetic correlations in Kondo's model \cite{Gu2012,Wei2012}. 
A recent investigation on Fe$_x$Pt$_{1-x}$ alloys also reported the enhancement in the SH effect near $T_C$ \cite{Ou2018}.
So far, the magnetic fluctuation at finite temperatures has been theoretically treated on a single-site level \cite{Guo2009,Gu2010a,Gu2010b} 
or using static approximations \cite{Kondo1962,Fert1972,Coleman1985,Fert1987,Gu2012}. 
When localized moments have long-range dynamical correlations near a magnetic instability, it is required to go beyond such a treatment 
(for example, see Refs.~\cite{Moriya1973,Hertz1976,Moriya1985,Millis1993}). 
This could open new pathways for novel spintronics. 

In this paper, we address the effect of such magnetic fluctuations onto the SH effect by calculating the SH conductivity of a model system 
in which conduction electrons are interacting with dynamically fluctuating local magnetic moments. 
We start from defining our model Hamiltonian and then identify two different mechanisms for the SH effect. 
The similarity and dissimilarity with the SH effect arising from impurity potential scattering or phonon scattering are discussed. 
The SH conductivity is computed using the Matsubara formalism by combining the self-consistent renormalization theory \cite{Moriya1985}. 
We show that the SH conductivity is enhanced at low temperatures when the system is in close vicinity to the ferromagnetic critical point at $T=0$. 
Possible realization of this effect in $4d$ or $5d$ metallic compounds is discussed.



{\it Model and formalism}.---%
To be specific, we consider the $s$-$d$ or $s$-$f$ Hamiltonian proposed by Kondo \cite{Kondo1962,supp}, 
$H = H_0 + H_K$ with $H_0=\sum_{\vect k, \nu} \varepsilon_{\vect k} a_{\vect k \nu}^\dag a_{\vect k \nu}$ and 
\begin{eqnarray}
&&\hspace{-1em} H_K =
-\frac{1}{N} \sum_{n}^{N_m} \sum_{\vect k, \vect k'} \sum_{\nu, \nu'} e^{i(\vect k' -\vect k) \cdot \vect R_n} 
a_{\vect k \nu}^\dag a_{\vect k' \nu'} \nonumber \\
&&\hspace{-1em} \times \biggl[ 2(\vect J_n \cdot \vect s_{\nu \nu'})
\bigl\{ {\cal F}_0 + 2 {\cal F}_1(\vect k \cdot \vect k') \bigr\} 
+ i {\cal F}_2 \vect J_n \cdot (\vect k' \times \vect k) \nonumber \\
&&\hspace{-1em} + i {\cal F}_3 \Bigl\{( \vect J_n \cdot \vect s_{\nu \nu'}) \bigl( \vect J_n \cdot (\vect k' \times \vect k) \bigr) 
+\bigl( \vect J_n \cdot (\vect k' \times \vect k) \bigr) (\vect J_n \cdot \vect s_{\nu \nu'}) \nonumber \\
&&\hspace{-1em} - \frac{2}{3} (\vect J_n \cdot \vect J_n) \bigl( \vect s_{\nu \nu'} \cdot (\vect k' \times \vect k) \bigr) \Bigr\} \biggr] . 
\label{eq:Kondo}
\end{eqnarray}
Here, $a_{\vect k \nu}^{(\dag)}$ is the annihilation (creation) operator of a conduction electron with momentum $\vect k$ and spin $\nu$, 
$\varepsilon_{\vect k} = \frac{\hbar^2 k^2}{2m}- \mu$ is the dispersion relation measured from the Fermi level $\mu$ 
with the carrier effective mass $m$, 
$\vect s_{\nu \nu'}=\frac{1}{2} \vecs \sigma_{\nu \nu'}$is  the conduction electron spin with $\vecs \sigma$ the Pauli matrices,
and  $N (N_m)$ is the total number of lattice sites (local moments). 
$\vect J_n$ is the local spin moment at position $\vect R_n$, when the SOC is weaker than the crystal field splitting and could be treated as a perturbation, 
or the local total angular momentum, when the SOC is strong so that  the total angular momentum is a constant of motion.  
Parameters ${\cal F}_l$ are related to $F_l$ defined in Ref.~\cite{Kondo1962} 
as discussed in the Supplemental Material \cite{supp}. 
In this work, we focus on three-dimensional systems. 
While the current analysis could be applied to other dimensions, lower-dimensional systems require more careful treatments. 

\begin{figure}
\begin{center}
\includegraphics[width=0.9\columnwidth, clip]{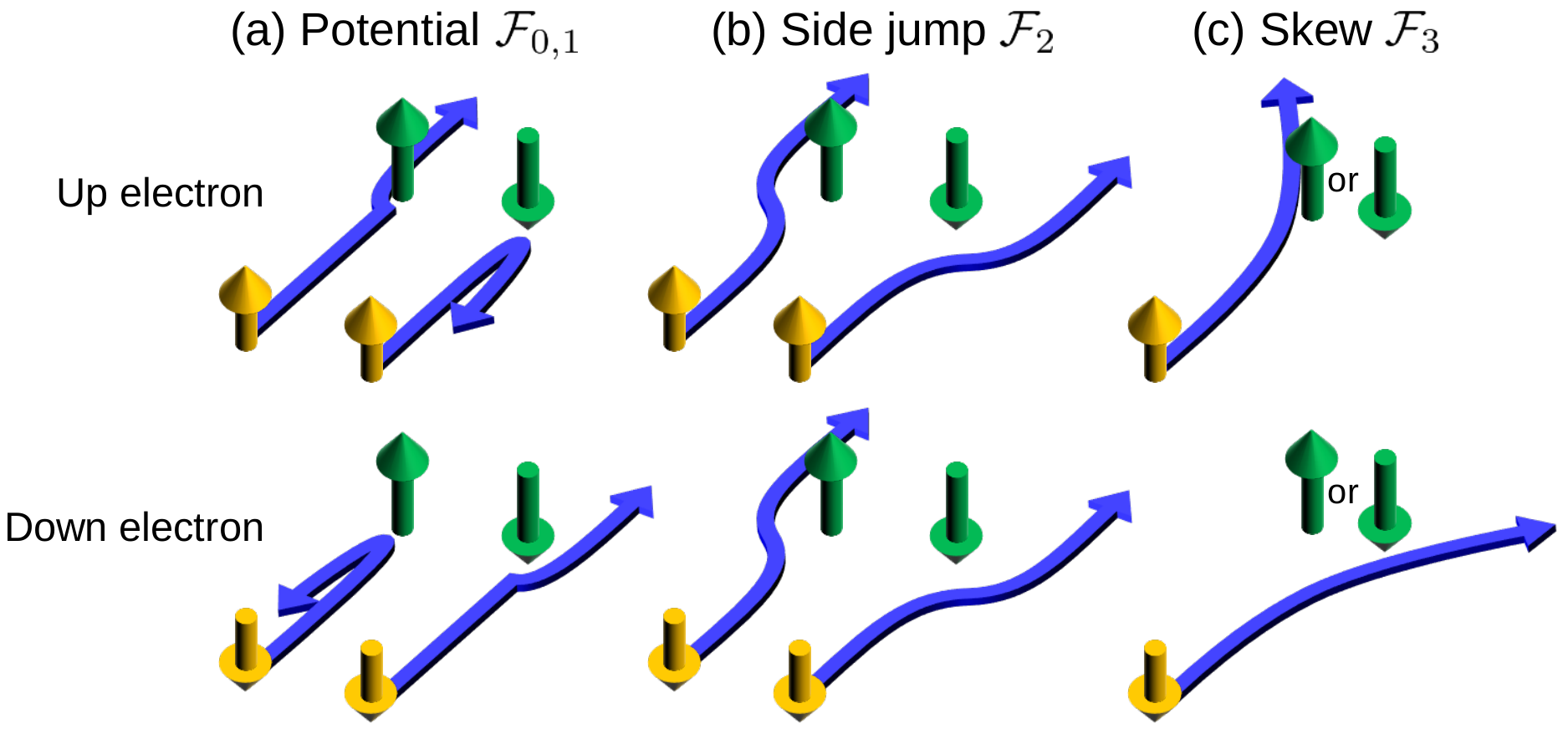}
\caption{Scattering processes involving  
(a) ${\cal F}_{0,1}$ terms, (b) ${\cal F}_2$ terms, and (c) ${\cal F}_3$terms. 
Yellow arrows indicate conduction electrons, and green arrows indicate local moments. 
In the ${\cal F}_{2 (3)}$ scattering processes, the electron deflection depends on the direction of the local moment (the electron spin), 
leading to the side-jump-type (skew-scattering-type) contribution to $\sigma_{SH}$. 
}
\label{fig:scatter}
\end{center}
\end{figure}

In Eq.~(\ref{eq:Kondo}), ${\cal F}_{0,1}$ terms correspond to the standard $s$-$d$ or $s$-$f$ exchange interaction, 
acting as the spin-dependent potential scattering as schematically shown in Fig.~\ref{fig:scatter}~(a). 
${\cal F}_{2,3}$ terms represent the exchange of angular momentum between a conduction electron and a local moment. 
These terms are odd (linear or cubic) order in $J_n$ and $s$ and induce the electron deflection depending on the direction of $\vect J_n$ or $\vect s$ 
as depicted in Figs.~\ref{fig:scatter}~(b) and ref{fig:scatter}~(c). 
As discussed below, the ${\cal F}_2$ term and the ${\cal F}_3$ term, respectively, generate 
the side-jump- and the skew-scattering-type contributions to the SH conductivity. 

In order to see the different types of contributions, 
we analyze the velocity operator, from which the charge current and the spin current operators are defined. 
Importantly, a side-jump-type contribution to the SH effect arises from the anomalous velocity as in the conventional SH effect. 
The velocity operator is defined by $\vect v = (i/\hbar) [H,\vect r]$. 
Among various terms, lowest order contributions to the spin Hall conductivity come from 
\begin{eqnarray}
\vect v  \!\!\!&=&\!\!\! \sum_{\vect k}\frac{\hbar \vect k}{m} a_{\vect k \nu}^\dag a_{\vect k \nu} 
- \frac{i }{\hbar N} \sum_{n} \sum_{\vect k, \vect k'} \sum_{\nu, \nu'} e^{i(\vect k'-\vect k) \cdot \vect R_n} 
\nonumber \\
\!\!\!&\times&\!\!\!  \bigl\{ {\cal F}_2 \vect J_n  
+ 2 {\cal F}_3  (\vect J_n \cdot \vect s_{\nu \nu'}) \vect J_n \bigr\} 
\times (\vect k'-\vect k) a_{\vect k \nu}^\dag a_{\vect k' \nu'}. 
%
%
\label{eq:velocity}
\end{eqnarray}
Here, a term involving ${\cal F}_1$ is neglected because it is proportional to $(\vect k + \vect k')$ and 
does not contribute to $\sigma_{SH}$ at the lowest order. 
The second terms involving ${\cal F}_{2,3}$ are the anomalous velocity. 
The charge current and the spin current are then given by using the velocity operator as 
$\vect j^c = -e \vect v$ and $\vect j^s = -e \{\frac{1}{N} \sum_{\vect k} s_{\nu \nu'}^z a_{\vect k \nu}^\dag a_{\vect k \nu'}, \vect v\}$, respectively. 
Note that $\vect j^c$ and $\vect j^s$ have the same dimension.

\begin{figure}
\begin{center}
\includegraphics[width=0.9\columnwidth, clip]{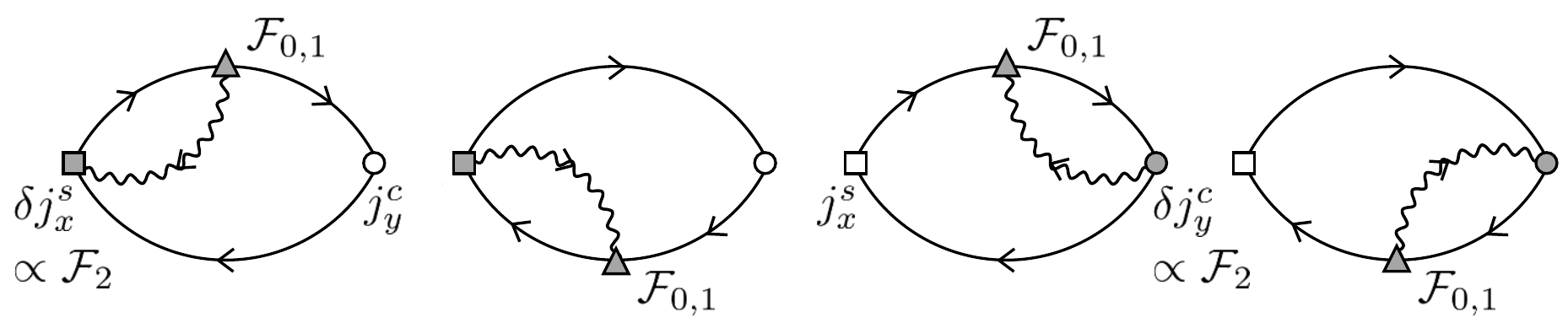}
\caption{Diagrammatic representation for the side-jump contribution. 
Solid (wavy) lines are the electron Green's functions (the spin fluctuation propagators). 
Squares (circles) are the spin (charge) current vertices, with filled symbols representing the velocity correction with ${\cal F}_2$, 
i.e., side jump. 
Filled triangles are the interaction vertices with ${\cal F}_{0,1}$. }
\label{fig:sidejump}
\end{center}
\end{figure}

Now, we consider the side-jump-type mechanism arising from the anomalous velocity in Eq.~(\ref{eq:velocity}) 
combined with the spin-dependent potential scattering ${\cal F}_{0,1}$ in Eq.~(\ref{eq:Kondo}). 
At this moment, one could notice some analogy between the current model and the previous ones utilizing the potential scattering $V_n$ 
\cite{Berger1970,Crepieux2001,Tse2006} as 
${\cal F}_{0,1} J_n s \leftrightarrow V_n$ and ${\cal F}_2 J_n \leftrightarrow \lambda^2 V_n s $, i.e., 
the spin $s$ dependence is switched from the anomalous velocity to the scattering term. 
Therefore, the second-order processes involving ${\cal F}_{0,1}$ and ${\cal F}_2$ terms could generate the side-jump-type contribution to the SH effect. 
The diagramatic representation of this side-jump-type contribution to the SH conductivity is presented in Fig.~\ref{fig:sidejump}. 
Note that this contribution is ${\cal O}({\cal F}_{0,1} {\cal F}_2 \langle J_n J_{n'} \rangle)$. 
If the  ${\cal F}_3$ term in the anomalous velocity is used, it would become 
${\cal O}({\cal F}_{0,1} {\cal F}_3 \langle J_n J^2_{n'} \rangle)$, odd order in the local moment. 
Such a contribution vanishes when the local moments have the TRS in a paramagnetic phase above magnetic transition temperature. 

How about the skew-scattering-type contribution? 
Unlike the side-jump-type contribution, the ${\cal F}_2$ does not contribute to $\sigma_{SH}$ arising from 
the third-order perturbation processes combined with ${\cal F}_{0,1}$ terms. 
This is because such processes are ${\cal O}({\cal F}^2_{0,1} {\cal F}_2 \langle J_n J_{n'} J_{n''} \rangle)$ and vanish by the TRS in the local moments. 
In fact, the skew-scattering-type contribution arises from the third-order processes involving ${\cal F}_{0,1}$ and ${\cal F}_3$ terms 
as ${\cal O}({\cal F}^2_{0,1} {\cal F}_3 \langle J_n J_{n'} J^2_{n''} \rangle)$. 
Therefore, such skew-scattering-type contributions are possible without introducing unharmonic (third-order) magnetic correlations, 
while it is second order in the spin fluctuation propagator ${\cal O}(D^2)$ as discussed below. 
This contrasts with the phonon skew scattering, where unharmonic phonon interactions are essential \cite{Gorini2015}. 

\begin{figure}
\begin{center}
\includegraphics[width=0.9\columnwidth, clip]{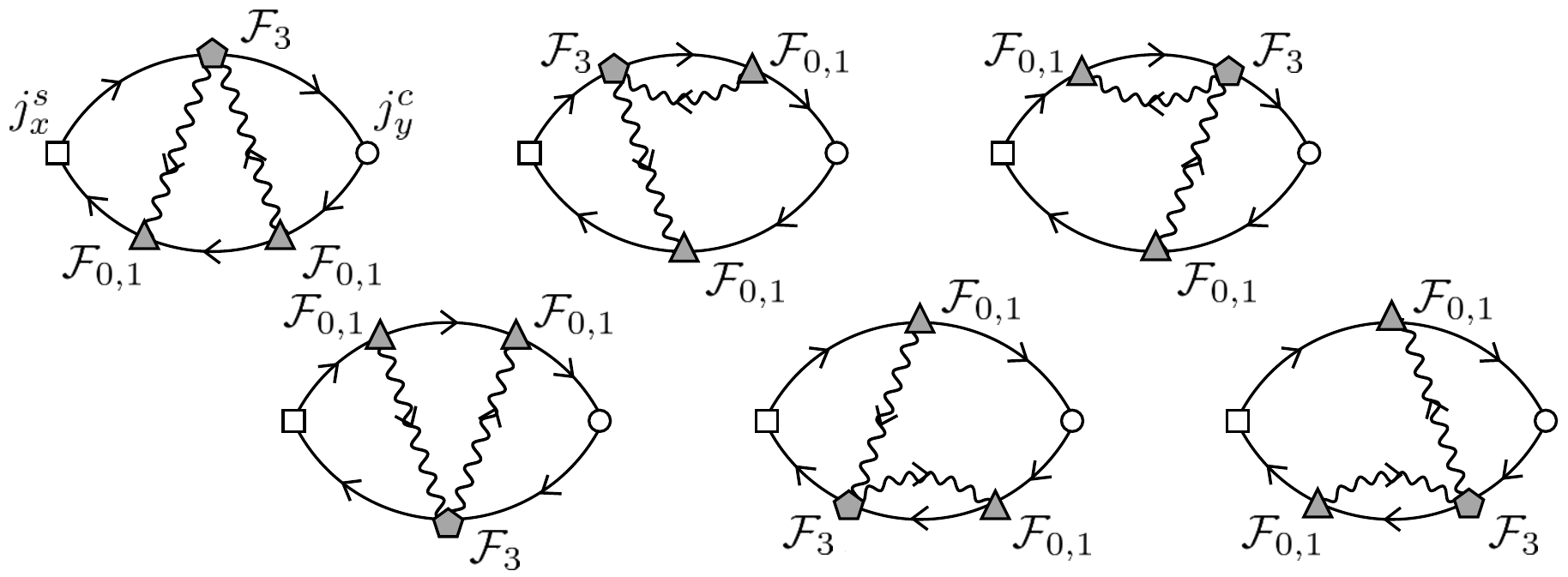}
\caption{Diagrammatic representation for the skew-scattering contribution. 
Filled pentagons are the interaction vertices with ${\cal F}_3$. 
The definitions of the other symbols or lines are the same as in Fig.~\ref{fig:sidejump}. }
\label{fig:skew}
\end{center}
\end{figure}

{\it Matsubara formalism and spin fluctuation}.---%
In what follows, we use the Matsubara formalism to compute the SH conductivity given by 
\begin{eqnarray}
\sigma_{SH}(i\Omega_l) = \frac{i}{i\Omega_l V} \int_0^{1/T} \hspace{-1em} d\tau e^{i \Omega_l \tau} 
\langle T_\tau j_x^s (\tau) j_y^c (0) \rangle, 
\label{eq:sigmaSH}
\end{eqnarray}
where $\Omega_l$ is the bosonic Matsubara frequency, and $V$ is the volume of the system. 
At the end of the analysis, $i \Omega_l$ is analytically continued to real frequency as $i\Omega_l \rightarrow \Omega + i 0^+$. 
We will then consider the dc limit, $\Omega \rightarrow 0$, to obtain $\sigma_{SH}$. 

%
This formalism allows one to treat conduction electrons coupled with dynamically fluctuating local moments $\vect J_n$. 
To describe the latter, we consider a generic Gaussian action given by 
$A_{Gauss} = \frac{1}{2}\sum_{\vect q, l}
D^{-1}_{\vect q}( i\omega_l) J_{\vect q}( i\omega_l) J_{-\vect q} (-i\omega_l)
$ 
with 
$D^{-1}_{\vect q} (i\omega_l) = \delta + A q^2 + |\omega_l|/\Gamma_q$. 
Here, $\omega_l = 2 l \pi T$ is the bosonic Matsubara frequency, 
and $A$ is introduced as a constant so that $A q^2$ has the unit of energy. 
$\delta$ is the distance from a ferromagnetically ordered state and is related to the magnetic correlation length as $\xi^2 \propto \delta^{-1}$. 
$J_{\vect q}( i\omega_l)$ is a space and imaginary-time $\tau$ Fourier transform of $J_{n}(\tau)$, where we made the $\tau$ dependence explicit. 
In principle, $\delta$ depends on temperature and is determined by solving self-consistent equations for a full model including non-Gaussian terms 
\cite{Moriya1973,Hertz1976,Moriya1985,Millis1993,Nagaosa1999}. 
$\Gamma_q$ represents the momentum-dependent damping. 
In clean metals close to the ferromagnetic instability, $\Gamma_q = \Gamma q$. 
When elastic scatting exists due to impurities or disorders, $q$ has a small cutoff $q_c \sim \ell^{-1} = 1/v_F \tau_c $ with 
$\ell$ being the mean free path of conduction electrons, $v_F = \hbar k_F/m$ the Fermi velocity, and $\tau_c$ the carrier lifetime. 
Therefore, the damping term at $q \alt q_c$ has to be replaced by $\Gamma q_c$ \cite{Lee1992}. 
With this propagator $D$, the spatial and temporal correlation of $\vect J_n$ is given by 
$\langle T_\tau J_{n} (\tau) J_{n'} (0) \rangle
= \frac{T}{N} \sum_{\vect q, l} e^{-i\omega_l \tau + i \vect q \cdot (\vect R_n - \vect R_{n'})} 
D_{\vect q} (i\omega_l)$. 
Theoretical analyses based on this model have been successful to explain many experimental results on itinerant magnets \cite{Moriya1985}. 

Because of the phase factor $ e^{ i \vect q \cdot (\vect R_n - \vect R_{n'})}$,  
the ferromagnetic fluctuation is essential for the SH effect. 
When the spin fluctuation has characteristic momentum $\vect Q \ne  \vect 0$, 
$ e^{ i \vect Q \cdot (\vect R_n - \vect R_{n'})}$ has destructive effects.

{\it Spin-Hall conductivity}.---%
With the above preparations, now we proceed to examine the SH conductivity. 
Based on the diagrammatic representations in Figs.~\ref{fig:sidejump} and \ref{fig:skew}, 
$\sigma_{SH}$ is expressed in terms of electron Green's function $G$ and the propagator of local magnetic moments $D$. 
The full expression is presented in Ref.~\cite{supp}. 

We carry out the Matsubara summations, the energy integrals and the momentum summations as detailed in Ref.~\cite{supp} to find 
\begin{equation}
\sigma^{side \, jump}_{SH} \approx \frac{2 e^2 n_m^2}{m} \, \tau_c \, I(T,\delta) \,
\biggl( \frac{1}{3} {\cal F}_0 k_F^2 - \frac{2}{5} {\cal F}_1 k_F^4 \biggr) {\cal F}_2 N_F 
\label{eq:sigmaSJ2}
\end{equation}
for the side-jump contribution and 
\begin{equation}
\sigma^{skew \, scat.}_{SH} \approx 
 \frac{4 e^2 \hbar n_m^3}{ m^2 }
\, \tau_c^2 \, I^2(T,\delta) \,
\Bigl( {\cal F}_0 + {\cal F}_1 k_F^2 \Bigr)^2 {\cal F}_3  \frac{2 k_F^4}{15} N_F
\label{eq:sigmaSS6}
\end{equation}
for the skew-scatting contribution. 
Here, 
$n_m=N_m/N$ is the concentration of local moments, 
and 
$N_F=m k_F/2 \pi^2 \hbar^2$ is the electron density of states per spin at the Fermi level. 
The function $I(T,\delta)$ defined in Ref.~\cite{supp} 
is the direct consequence of the coupling between conduction electrons and the dynamical spin fluctuation. 
There are a number of limiting cases where the analytic form of $I(T,\delta)$ is available. 
For clean systems ($\Gamma_q = \Gamma q$, i.e., no momentum cutoff) at low temperatures, 
where $\delta + A (a T/\hbar v_F)^2 \ll \hbar v_F/\Gamma$ is satisfied, 
$I(T,\delta) \approx \frac{1}{8 \pi \delta} (a T/\hbar v_F)^3$ 
with $a$ being the lattice constant. 
When the system is on the quantum critical point for the ferromagnetic ordering, 
$\delta$ is scaled as $\delta \propto T^{4/3}$~\cite{Moriya1985}. 
Thus, $I(T,\delta) \propto T^{5/3}$ is expected. 
For clean systems at high temperatures, where $\delta + A (a T/\hbar v_F)^2 \gg \hbar v_F/\Gamma$ is satisfied,  
$I(T,\delta) \approx \frac{\hbar v_F}{4 \pi^2 \Gamma \delta^2} (a T/\hbar v_F)^3$. 
At such high temperatures, $\delta$ is linearly dependent on $T$ \cite{Moriya1985,Ueda1975}. 
Therefore, one expects $I(T,\delta) \propto T$. 
Similar analyses are possible for dirty systems, where $\Gamma_q$ has a small momentum cutoff. 
In this case, one expects $I(T,\delta) \propto T$ at both low temperatures and high temperatures 
(see Ref.~\cite{supp} for details). 

\begin{figure}
\begin{center}
\includegraphics[width=0.9\columnwidth, clip]{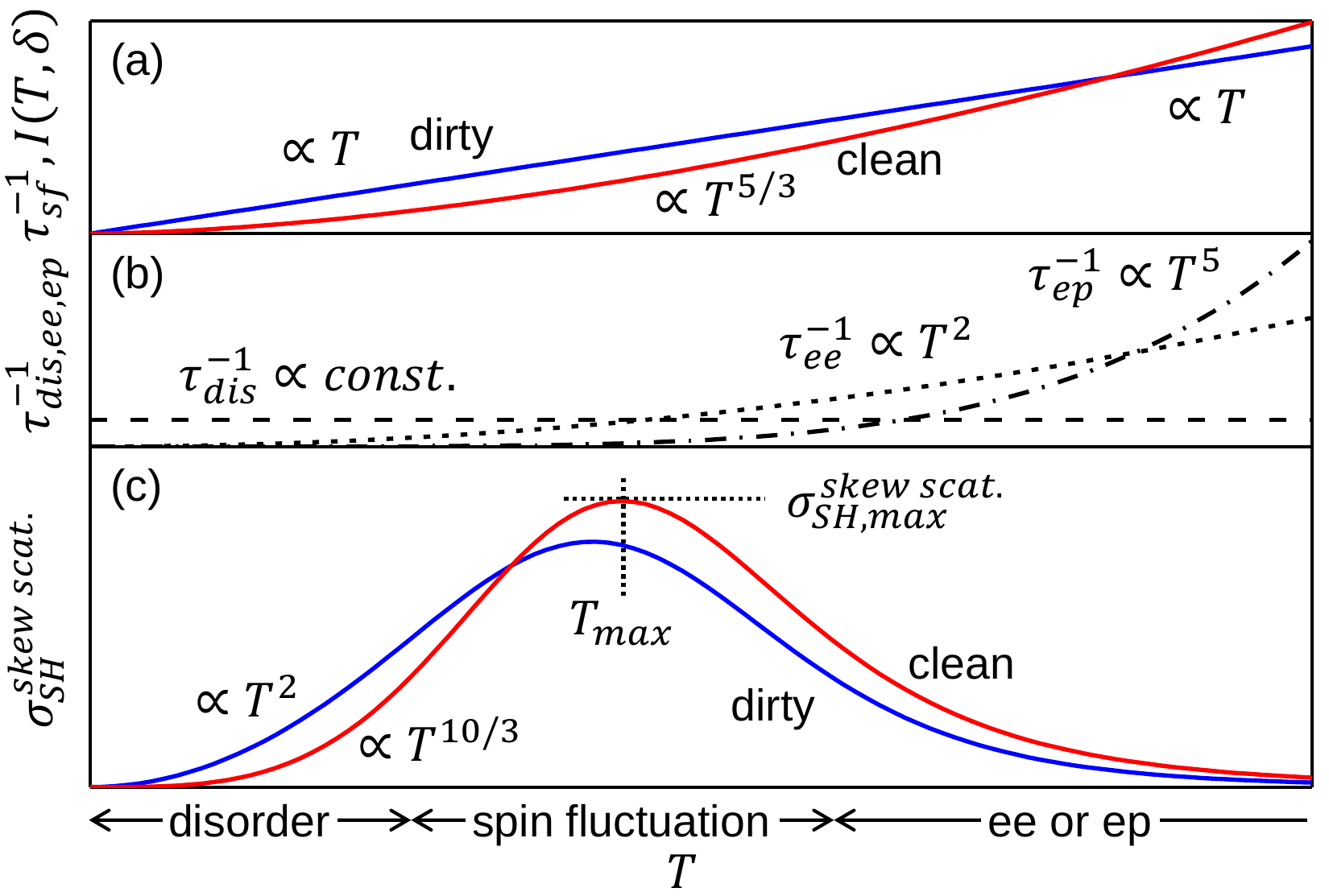}
\caption{Schematic temperature dependence of (a) $\tau_{sf}^{-1}$ and $I(T,\delta)$, 
(b) $\tau_{dis}^{-1}$ (dashed line), $\tau_{ee}^{-1}$ (dotted line), and $\tau_{ep}^{-1}$ (dash-dotted line), 
and (c) $\sigma_{SH}^{skew scat.}$. 
Red lines and blue lines correspond to the clean system and the dirty system, respectively.
At a low (intermediate, high) temperature regime, $\tau_c^{-1}$ is dominated by $\tau_{dis}^{-1}$ ($\tau_{sf}^{-1}$, $\tau_{ee}^{-1}$ or $\tau_{ep}^{-1}$), 
creating $\sigma_{SH, max}^{skew \, scat.}$ at $T_{max}$.}
\label{fig:Tdep}
\end{center}
\end{figure}

In addition to $I(T,\delta)$, the temperature dependence of $\sigma_{SH}$ is induced by the carrier lifetime $\tau_c$. 
This quantity comes from several different contributions as 
\begin{eqnarray}
\tau_c^{-1} = \tau_{sf}^{-1}+\tau^{-1}_{ee} + \tau^{-1}_{ep}+ \tau_{dis}^{-1} + \ldots
\end{eqnarray}
Here, 
$\tau_{sf}^{-1}$ is from the scattering due to the spin fluctuation. 
Using $H_K$ and the same level of approximation, 
$\tau_{sf}^{-1}$ is given by $\tau_{sf}^{-1} \approx \frac{2 n_m^2}{\hbar} I(T,\delta) ({\cal F}_0+2 {\cal F}_1  k_F^2 )^2$ \cite{supp}. 
$\tau_{sf}^{-1}$ and $I(T,\delta)$ have the same $T$ dependence as schematically shown in Fig.~\ref{fig:Tdep}~(a). 
$\tau_{ee}^{-1}$ and $\tau_{ep}^{-1}$ are from the electron-electron interactions and the electron-phonon interactions, respectively.  
Their leading $T$ dependence is given by $\tau_{ee}^{-1} \approx \tau_{ee,0}^{-1}(T/T_F)^2$ \cite{Baber1937} 
and $\tau_{ep}^{-1} \approx \tau_{ep,0}^{-1}(T/T_D)^5$ \cite{Bloch1930,Ziman1960}, 
where $T_{F(D)}$ is the Fermi (Debye) temperature. 
$\tau_{dis}^{-1}$ is from the disorder effects, and its $T$ dependence is expected to be small. 
Figure~\ref{fig:Tdep}~(b) summarizes the $T$ dependence of $\tau^{-1}_{dis,ee,ep}$.

The overall $T$ dependence of $\sigma_{SH}$ is determined by the combination of  $I(T,\delta)$ and $\tau_c$. 
The strong enhancement is thus expected at the ferromagnetic critical point, where the magnetic correlation length $\xi \propto \delta^{-1/2}$ diverges as $T^{-2/3}$.  
This results in $\tau_{sf}^{-1}$ and hence the electrical resistivity $\sigma_c^{-1}$ scaled as $T^{5/3}$ \cite{Ueda1975}. 
%
Since $\tau_{sf}^{-1} \propto I(T,\delta)$, $\sigma^{side \, jump}_{SH}$ and $\sigma^{skew \, scat.}_{SH}$ are expected to be maximized 
when the spin fluctuation dominates $\tau_c$ as  
\begin{equation}
\sigma^{side \, jump}_{SH, max} \approx  
\frac{e^2\hbar}{m} \frac{{\cal F}_2 k_F^2 }{3 {\cal F}_0} N_F
\label{eq:sigmaSJ3}
\end{equation}
and 
\begin{equation}
\sigma^{skew\, scatt.}_{SH, max} 
\approx 
\frac{e^2 \hbar^3}{m^2 n_m} \frac{2 {\cal F}_3  k_F^4}{15 {\cal F}_0^2} N_F,
\label{eq:sigmaSS7}
\end{equation}
respectively, 
at low but nonzero temperature $T_{max}$. 
This $T_{max}$ is approximately given by $T_F (5 \tau_{ee,0}/\tau_{dis})^{1/2}$ when $T_F \ll T_D$ or 
$T_D (\tau_{ep,0}/ 2 \tau_{dis})^{1/5}$ when $T_F \gg T_D$. 
%
As the temperature is lowered to zero, $\sigma_{SH}$ goes to zero as $\sigma^{side \, jump}_{SH} \propto \tau_{dis} I (T,\delta) \propto T^{5/3}$ and 
 $\sigma^{skew \, scat.}_{SH} \propto \tau_{dis}^2 I^2(T,\delta) \propto T^{10/3}$
because of the nonzero $\tau_{dis}^{-1}$, 
and the residual SH conductivity is due to disorders or impurities. 
At higher temperatures, 
the carrier lifetime is suppressed by the electron-electron or electron-phonon interaction, and therefore 
$\sigma_{SH}$ is decreased. 
The overall $T$ dependence of $\sigma_{SH}^{skew \, scat.}$ is schematically shown in Fig.~\ref{fig:Tdep}~(c). 

In dirty systems, $\Gamma_q$ involves a small cutoff momentum. 
Because $\tau_{dis}$ is dominant, we expect $\sigma^{side \, jump}_{SH} \propto T$ 
and $\sigma^{skew \, scat.}_{SH} \propto T^2$ at low temperatures as discussed in Ref.~\cite{supp}.
When the temperature is increased above  $T \sim \min\{ T_F,T_D\}$, 
$\sigma_{SH}$ decreases with $T$ because $\tau_c$ is suppressed. 
Thus, $\sigma_{SH}$ is expected to be maximized at around $T_{max}$ as discussed for clean systems, 
yet the maximum value depends explicitly on $\tau_c$'s.  
In fact, the enhancement in $\tau_{sf,ee,ep}^{-1}$ with increasing $T$ 
always induces a momentum cutoff in the damping term $\Gamma_q$ at high temperatures. 
Therefore, we expect that clean systems and dirty systems behave similarly at high temperatures,
i.e., $\sigma_{SH}^{side \, jump} \propto \tau_c T$ and 
$\sigma_{SH}^{skew \, scat.} \propto \tau_c^2 T^2$.

{\it Discussion}.---%
How realistic is the current spin fluctuation mechanism? 
Here, we provide rough estimations of $\sigma_{SH,max}^{side \, jump}$ and $\sigma^{skew \, scat.}_{SH, max}$. 
According to a free electron model, ${\cal F}_0$ is expected to be $\sim 0.1$~eV for both transition metal and actinide compounds \cite{Kasuya1959}. 
(In Ref.~\cite{Kasuya1959}, $J_0$, corresponding to ${\cal F}_0$ in this study, was estimated to be $0.7\times 10^{-12}$~erg for the $s$-$d$ interaction in Mn and 
$2.5 \times 10^{-13}$~erg for the $s$-$f$ interaction in Gd.) 
Since ${\cal F}_{2,3} k_F^2$ involve the integral of higher-order spherical Bessel functions, $j_{1,3}$, i.e., $p$-wave scattering, 
than ${\cal F}_0$, $j_0$, i.e., $s$-wave scattering \cite{Kondo1962}, 
${\cal F}_{2(3)} k_F^2$ would be an order (two orders) of magnitude smaller than ${\cal F}_0$. 
Therefore, taking a rough estimation ${\cal F}_2 k_F^2 \sim 0.01$~eV, ${\cal F}_3 k_F^2 \sim 0.001$~eV and 
typical values of $k_F/\pi  \sim 10^9~{\rm m}^{-1}$ and $\frac{\hbar^2 k_F^2}{2m}=\mu \sim 10$~eV \cite{Martin} for $s$ electrons in metallic compounds, 
optimistic estimations are  $\sigma^{side \, jump}_{SH, max} \sim 10^{3}~\Omega^{-1} {\rm m}^{-1}$ and 
$\sigma^{skew \, scat.}_{SH, max} \sim 10^{5}~\Omega^{-1} {\rm m}^{-1}$. 
The difference in magnitude between  $\sigma^{side \, jump}_{SH, max}$ and  $\sigma^{skew \, scat.}_{SH, max}$ comes from 
the small factor ${\cal F}_2/{\cal F}_0$ in $\sigma^{side \, jump}_{SH, max}$ and the large factor $\mu/{\cal F}_0$ in  $\sigma^{skew \, scat.}_{SH, max}$. 
Thus, $\sigma^{skew \, scat.}_{SH, max}$ could be comparable to the largest $\sigma_{SH}$ reported so far \cite{Hoffmann2013}.

Could there be systems that show the SH effect by the proposed mechanisms? 
The crucial ingredients are the coupling between conduction electrons and localized but not ordered magnetic moments. 
Suitable candidate materials would be $4d$ or $5d$ metallic compounds with partially filled $d$ shells, such as  Ir, Pt, W and Re. 
Because of the large SOC than $3d$ compounds, 
the intrinsic mechanism could contribute to the SH effect. 
One route to enhance $\sigma_{SH}$ further is doping with magnetic $3d$ transition metal elements to enhance the ferromagnetic spin fluctuation. 
It would be possible to distinguish between the intrinsic mechanism and the extrinsic mechanisms discussed in this work 
by comparing crystalline samples and disordered samples such as metallic glasses. 
In fact, metallic glasses might be a good choice in trying to enhance the SH angle  $\Theta_{SH}$. 
Since the carrier lifetime in metallic glasses is dominated by the structure factor, 
the temperature dependence of $\tau_c \sim \tau_{dis}$ is small  \cite{Ziman1961,Ziman1967}. 
Using the same formalism, the longitudinal charge conductivity is given by $\sigma_c = 2e^2 \tau_c k_F^3/3m \pi^2$. 
Therefore, $\Theta_{SH} = \sigma_{SH}/\sigma_c$ is more sensitive to the spin fluctuation contribution than $\sigma_{SH}$ itself. 
Since $\sigma_{SH}^{skew \, scat.}$ is dominant, the spin fluctuation contribution $I(T,\delta)$ could be extracted from $\sigma_{SH}/\sigma_c^2$. 
%
Recently, Ou {\it et al.} reported very large $\Theta_{SH}>0.34$ in Fe$_x$Pt$_{1-x}$ alloys near $T_C$ \cite{Ou2018}. 
While the detailed analyses remain to be carried out, 
with the typical conductivity in their sample $\sigma_c \sim 10^{6}~ \Omega^{-1} {\rm m}^{-1}$ and 
our theoretical $\sigma^{skew\,  scat.}_{SH, max} \sim 10^{5}~\Omega^{-1} {\rm m}^{-1}$, 
 $\Theta_{SH}$ is estimated to be $\sim 0.1$, that is comparable to this report. 

To summarize, we investigated the effect of fluctuating magnetic moments on the spin Hall effect in metallic systems. 
We employed the microscopic model developed by Kondo for the coupling between conduction electrons and localized moments \cite{Kondo1962} and 
analyzed the fluctuation of local moments using the self-consistent renormalization theory by Moriya \cite{Moriya1985}. 
As in the conventional spin Hall effect due to the impurity scattering, a side-jump-type mechanism and a skew-scattering-type mechanism appear. 
Because of the dynamical spin fluctuation, the spin Hall conductivity has a nontrivial temperature dependence, 
leading to the enhancement at nonzero temperatures near the ferromagnetic instability.  
The skew scattering mechanism we proposed could generate a sizable spin Hall effect.

The research by S.O. and T.E. was supported by the  U.S. Department of Energy, Office of Science, Basic Energy Sciences, Materials Sciences and Engineering Division. 
N.N. was supported by JST CREST Grant No. JPMJCR1874 and JPMJCR16F1, Japan, and JSPS KAKENHI Grants No. 18H03676 and No. 26103006.
%
%


\onecolumngrid

\begin{center}
{\large \bf Supplementary material: Critical spin fluctuation mechanism for the spin Hall effect}\\
\vspace{1em}
Satoshi Okamoto,$^1$ Takeshi Egami,$^{1,2,3}$ and Naoto Nagaosa$^{4,5}$\\
\vspace{0.5em}
{\small \it $^1$Materials Science and Technology Division, Oak Ridge National Laboratory, Oak Ridge, Tennessee 37831, USA}\\
{\small \it $^2$Department of Materials Science and Engineering, The University of Tennessee, Knoxville, Tennessee 37996, USA}\\
{\small \it $^3$Department of Physics and Astronomy, The University of Tennessee, Knoxville, Tennessee 37996, USA}\\
{\small \it $^4$Department of Applied Physics, The University of Tokyo, Bunkyo-ku, Tokyo 113-8656, Japan}\\
{\small \it $^5$RIKEN Center for Emergent Matter Science (CEMS), Wako, Saitama 351-0198, Japan}\\
\end{center}

\renewcommand{\thetable}{S\Roman{table}}
\renewcommand{\thefigure}{S\arabic{figure}}
\renewcommand{\thesubsection}{S\arabic{subsection}}
\renewcommand{\thesubsubsection}{S\arabic{subsection}.\arabic{subsubsection}}
\renewcommand{\theequation}{S\arabic{equation}}

\setcounter{secnumdepth}{3}

\setcounter{equation}{0}
\setcounter{figure}{0}

\subsection{Relation between ${\cal F}_l$ and $F_l$ in Ref.~\cite{SKondo1962}}

This section provides the relation between ${\cal F}_l$ appearing in Eq.~(1) and $F_l$ defined in Ref.~\cite{SKondo1962}. 
To make this relation transparent, we express the original Hamiltonians derived in Ref.~\cite{SKondo1962} 
using a more tractable form as  Eq.~(1).  

First, we consider a weak spin-orbit coupling (SOC) case, 
where the SOC strength $\lambda$ is smaller than the crystal field splitting $\Delta$ and, therefore, the SOC can be treated as a perturbation. 
For this case, the local exchange term given in Eq.~(2.33) is rewritten as 
\begin{eqnarray}
H_K \!\!\!&=&\!\!\!
-\frac{1}{N} \sum_{n}^{N_m} \sum_{\vect k, \vect k'} \sum_{\nu, \nu'} e^{i(\vect k' -\vect k) \cdot \vect R_n} 
a_{\vect k \nu}^\dag a_{\vect k' \nu'} 
\biggl[ 2(\vect S_n \cdot \vect s_{\nu \nu'})
\bigl\{ F_0 + 2 F_1(\vecs \kappa \cdot \vecs \kappa') \bigr\} 
+ i  \Lambda_1 F_2 \vect S_n \cdot (\vecs \kappa' \times \vecs \kappa) \nonumber \\
&+&\!\!\! i 2 \Lambda_1 c_2 F_2 \Bigl\{( \vect S_n \cdot \vect s_{\nu \nu'}) \bigl( \vect S_n \cdot (\vecs \kappa' \times \vecs \kappa) \bigr) 
+\bigl( \vect S_n \cdot (\vecs \kappa' \times \vecs \kappa) \bigr) (\vect S_n \cdot \vect s_{\nu \nu'}) 
- \frac{2}{3} (\vect S_n \cdot \vect S_n) \bigl( \vect s_{\nu \nu'} \cdot (\vecs \kappa' \times \vecs \kappa) \bigr) \Bigr\} \biggr] .  
\label{eq:Kondo2}
\end{eqnarray}
Here, $\vect S_n$ is a local spin moment at site $\vect R_n$, 
$\vecs \kappa$ and $\vecs \kappa'$ are the unit vectors in the directions of $\vect k$ and $\vect k'$, respectively. 
Parameters $F_l$ are exchange interactions between local $d$ or $f$ orbitals and the conduction electron 
as defined in Eqs.~(2.15--18). 
$\Lambda_1$ is a dimensionless parameter roughly proportional to $\lambda/\Delta$ 
as defined in Eq.~(2.34) or (2,39) in Ref.~\cite{SKondo1962}. 
$c_2$ is also a dimensionless parameter ${\cal O}(1/10)$--${\cal O}(1)$ depending on the electron configuration of a magnetic site
as summarized in Table~I in Ref.~\cite{SKondo1962}. 
We neglected terms that are higher order in $\lambda$. 

We next consider a strong SOC case, where $\lambda \gg \Delta$ and the total angular momentum, a sum of spin momentum and angular momentum, 
is a constant of motion.  
The corresponding Eq.~(2.45) is rewritten as 
\begin{eqnarray}
H_K \!\!\!&=&\!\!\!
-\frac{1}{N} \sum_{n}^{N_m} \sum_{\vect k, \vect k'} \sum_{\nu, \nu'} e^{i(\vect k' -\vect k) \cdot \vect R_n} 
a_{\vect k \nu}^\dag a_{\vect k' \nu'} 
\biggl[ 2 (g_J-1)  (\vect J_n \cdot \vect s_{\nu \nu'})
\bigl\{ F_0 + 2 F_1(\vecs \kappa \cdot \vecs \kappa') \bigr\} 
+ \frac{i}{2} (2-g_J) F_2 \vect J_n \cdot (\vecs \kappa' \times \vecs \kappa) \nonumber \\
&+&\!\!\! i d_3 F_3 \Bigl\{( \vect J_n \cdot \vect s_{\nu \nu'}) \bigl( \vect J_n \cdot (\vecs \kappa' \times \vecs \kappa) \bigr) 
+\bigl( \vect J_n \cdot (\vecs \kappa' \times \vecs \kappa) \bigr) (\vect J_n \cdot \vect s_{\nu \nu'}) 
- \frac{2}{3} (\vect J_n \cdot \vect J_n) \bigl( \vect s_{\nu \nu'} \cdot (\vecs \kappa' \times \vecs \kappa) \bigr) \Bigr\} \biggr] . 
\label{eq:Kondo3}
\end{eqnarray}
Here, $\vect J_n$ is a local total angular momentum at site $\vect R_n$, and $g_J$ is the Land{\'e} $g$ factor. 
 $d_3$ is a dimensionless parameter of ${\cal O}(1/10)$ -- ${\cal O}(1)$ depending on the electron configuration of a magnetic site. 
 This parameter is defined in Eq.~(2.48) in Ref.~\cite{SKondo1962}. 
We neglected terms which contain quadrupole moments because those terms do not contributed to the spin Hall (SH) effect. 

In Eqs.~(\ref{eq:Kondo2}) and (\ref{eq:Kondo3}), $F_{1,2,3}$ terms contain $\vecs \kappa$ and $\vecs \kappa'$, instead of $\vect k$ and $\vect k'$. 
We rewrite these terms by replacing $\vecs \kappa$ and $\vecs \kappa'$ by $\vect k/k_F$ and $\vect k'/k_F$, respectively. 
When $k$ and $k'$ are away from $k_F$, one has to consider higher order terms with respect to $k$ and $k'$.  
These terms are expected to produce higher-order corrections with respect to temperature $T$ in our results.  
However, such corrections are expected to be small because only $|\vect k| \sim k_F$ contributes in our analyses. 

After this replacement, the correspondence between Eq.~(1) in the main text and Eq.~(\ref{eq:Kondo2}) or (\ref{eq:Kondo3}) 
is clearer. 
For a weak SOC case, 
$\vect J_n \Leftrightarrow \vect S_n$, ${\cal F}_0 \Leftrightarrow F_0$, ${\cal F}_1 \Leftrightarrow \frac{1}{k_F^2} F_1$, 
${\cal F}_2 \Leftrightarrow \frac{1}{k_F^2} \Lambda_1 F_2$, and ${\cal F}_3 \Leftrightarrow \frac{2}{k_F^2} c_2 \Lambda_1 F_2$. 
For a strong SOC case, 
$\vect J_n \Leftrightarrow \vect J_n$
${\cal F}_0 \Leftrightarrow (g_J-1 ) F_0$, ${\cal F}_1 \Leftrightarrow \frac{1}{k_F^2} (g_J-1 ) F_1$, 
${\cal F}_2  \Leftrightarrow \frac{1}{k_F^2} (2-g_J)F_2/2$, and ${\cal F}_3  \Leftrightarrow \frac{1}{k_F^2} d_3 F_3$. 
Thus, our model Eq.~(1) unifies weak SOC and strong SOC cases.

\subsection{Side jump}

The SH conductivity by the side-jump-type mechanism as diagramatically shown in Fig.~2 
is expressed in terms of the electron Green's function and the propagator of the spin fluctuation as 
\begin{eqnarray}
\sigma^{side \, jump}_{SH} (i\Omega_l) 
\!\!\!&=&\!\!\! \frac{1}{i \Omega_l} \frac{2e^2}{m} \frac{T^2}{V N^3} \sum_{l, l'} \sum_{n,n'} \sum_{\vect k, \vect k'} 
\Bigl\{{\cal F}_0 k_x^2 - 2 {\cal F}_1 k_x^2 \bigl( k'_x \bigr)^2 \Bigr\} {\cal F}_2 \nonumber \\
&\times&\!\!\! G_{\vect k} (i\varepsilon_l) G_{\vect k}(i\varepsilon_l + i \hbar \Omega_l) 
\bigl\{ G_{\vect k'}(i\varepsilon_{l'}+i \hbar \Omega_l) - G_{\vect k'}(i\varepsilon_{l'}) \bigr\} \nonumber \\
&\times& \!\!\! D_{\vect k'-\vect k}(i\varepsilon_{l'} -i\varepsilon_l)  e^{i (\vect k' -\vect k)\cdot (\vect R_n -\vect R_{n'})}, 
\label{eq:sigmaSH0}
\end{eqnarray}
where, $G_{\vect k}(i \varepsilon_l)=\{i \varepsilon_l-\varepsilon_{\vect k} - \Sigma_{\vect k} (i \varepsilon_l)\}^{-1}$ 
is the electron Matsubara Green's function, 
with the fermionic Matsubara frequency $\varepsilon_l = (2l+1) \pi T$. 
Planck constant $\hbar$ is included explicitly in front of the Matsubara frequency $\Omega_l$.

After carrying out the Matsubara summation, and taking the limit of $i\Omega_l \rightarrow 0$, one obtains 
\begin{eqnarray}
\sigma^{side \, jump}_{SH}\!\!\!&=&\!\!\! \frac{2e^2 \hbar}{\pi m V N^3} \! \int \!\! d\varepsilon d \omega \sum_{n,n'} \sum_{\vect k, \vect k'} 
\Bigl\{{\cal F}_0 k_x^2 - 2 {\cal F}_1 k_x^2 \bigl( k'_x \bigr)^2 \Bigr\} {\cal F}_2 
 B_{\vect k' -\vect k}(\omega) \Bigl\{ b(\omega) + f \bigl(\varepsilon_{\vect k'} \bigr) \Bigr\}  \nonumber \\
&\times&\!\!\! \Bigl[\partial_\varepsilon f(\varepsilon) 
\bigl| G_{\vect k}^R(\varepsilon) \bigr|^2
\Im \bigl\{ G_{\vect k'}^R(\varepsilon+\omega) \bigr\} 
+ f(\varepsilon) \Im \bigl\{ G_{\vect k}^R(\varepsilon) G_{\vect k}^R(\varepsilon) 
\partial_\varepsilon G_{\vect k'}^R(\varepsilon+\omega) \bigr\} \Bigr] 
e^{i (\vect k' -\vect k)\cdot (\vect R_n -\vect R_{n'})}. 
\label{eq:sigmaSJ1}
\end{eqnarray}
$f(\varepsilon)$ and $b(\omega)$ are the Fermi distribution function and the Bose distribution function, respectively. 
$G_{\vect k}^{R,A}(\varepsilon) = G_{\vect k}(i \varepsilon_l \rightarrow \varepsilon \pm i \hbar /2 \tau_c)$ 
are the retarded and advanced Green's function, respectively. 
Here, the self-energy is assumed to be independent of $\varepsilon$, and $\tau_c$ is the carrier lifetime. 
$B_{\vect q}(\omega)$ is the spectral function of the $J$ propagator 
given by $B_{\vect q}(\omega)= -\frac{1}{\pi} \Im D_{\vect q}(i\omega_l \rightarrow \omega +i0^+) 
= \frac{1}{\pi} \frac{\omega/\Gamma_q}{( \delta + A q^2 )^2 +(\omega/\Gamma_q)^2}$. 

The first term in the square bracket of Eq.~(\ref{eq:sigmaSJ1}) is proportional to $\partial_\varepsilon f(\varepsilon) \approx - \delta(\varepsilon)$, the so-called Fermi surface term, 
while the second term is proportional to $f(\varepsilon)$, the so-called Fermi sea term. 
In principle, two terms contribute, but it can be shown that the contribution from the second term, the Fermi sea term, is small. 
Thus, we focus on the first contribution.

We use the following approximations considering the small self-energy $\Sigma_{\vect k} (\varepsilon) = i \hbar/2 \tau_c$: 
$|G^R_{\vect k} (\varepsilon)|^2 \approx (2 \pi \tau_c /\hbar) \delta(\varepsilon - \varepsilon_{\vect k}) $ and 
$\Im G^R_{\vect k} (\varepsilon) \approx -\pi \delta(\varepsilon - \varepsilon_{\vect k}) $. 
Performing the $\varepsilon$ and $\omega$ integrals in Eq.~(\ref{eq:sigmaSJ1}), one obtains
\begin{eqnarray}
\sigma^{side \, jump}_{SH} 
\!\!\!&\approx&\!\!\! \frac{2 e^2 \pi \tau_c}{m V N^3} \sum_{n,n'} \sum_{\vect k, \vect k'} \delta(\varepsilon_{\vect k})
\Bigl\{{\cal F}_0 k_x^2 - 2 {\cal F}_1 k_x^2 \bigl( k'_x \bigr)^2 \Bigr\} {\cal F}_2  
B_{\vect k' - \vect k} \bigl(\varepsilon_{\vect k'} \bigr) \Bigl\{ b\bigl(\varepsilon_{\vect k'}\bigr) + f \bigl( \varepsilon_{\vect k'} \bigr)\Bigr\} 
e^{i (\vect k' -\vect k)\cdot (\vect R_n -\vect R_{n'})}  .
\end{eqnarray}

Noticing that $B_{\vect k'-\vect k}$ is dominated by small $|\vect k'-\vect k|$ regions, 
we replace $\vect k'$ by $\vect k + \vect q$ and $e^{i \vect q\cdot (\vect R_n -\vect R_{n'})}$ by $1$. 
Then, $\varepsilon_{\vect k'}$ is approximated as 
$\varepsilon_{\vect k'}=\frac{\hbar^2}{2m}|\vect k + \vect q|^2 -\mu \approx \varepsilon_{\vect k} -\mu + \hbar \vect v_F \cdot \vect q$ 
near the Fermi level, with $\vect v_F$ being the Fermi velocity parallel to $\vect k$. 
This leads to 
\begin{eqnarray}
\hspace{-1em}\sigma^{side \, jump}_{SH} 
\!\!\!&\approx&\!\!\! \frac{2 e^2 \pi \tau_c}{m V N} n_m^2 
\sum_{\vect k, \vect q} \delta(\varepsilon_{\vect k})
\Bigl\{{\cal F}_0 k_x^2 - 2 {\cal F}_1 k_x^2 (k_x+q_x)^2 \Bigr\} {\cal F}_2 
B_{\vect q} \bigl( \hbar \vect v_F \cdot \vect q \bigr) 
\Bigl\{ b\bigl( \hbar \vect v_F \cdot \vect q\bigr) + f \bigl( \hbar \vect v_F \cdot \vect q \bigr)\Bigr\} .
\label{eq:sigmaSJS1}
\end{eqnarray}
Here, $n_m=N_m/N$ is the concentration of local moments. 
By neglecting small corrections coming from $q_x^2$, the $\vect q$ integral is summarized into the following function, 
\begin{eqnarray}
I(T,\delta) \equiv
\frac{\pi}{N} \sum_{\vect q} 
B_{\vect q} \bigl( \hbar \vect v_F \cdot \vect q \bigr) 
\Bigl\{ b\bigl( \hbar \vect v_F \cdot \vect q\bigr) + f \bigl( \hbar \vect v_F \cdot \vect q \bigr)\Bigr\} .
\label{eq:F}
\end{eqnarray}
Combining Eqs.~(\ref{eq:sigmaSJS1}) and (\ref{eq:F}), one arrives at Eq.~(4). 

\subsection{Skew scattering}

Using Matsubara Green's functions for conduction electrons and the spin fluctuation, 
the SH conductivity due to the skew-type scattering is expressed as 
\begin{eqnarray}
\sigma^{skew \, scat.}_{SH} (i\Omega_l) \!\!\!&=&\!\!\! \frac{1}{i \Omega_l} \frac{e^2 \hbar^2}{m^2} \frac{T^3}{V N^5} \sum_{l, l', l''} \sum_{n,n',n''} \sum_{\vect k, \vect k', \vect k''} 
\nonumber \\
&\times&\!\!\! G_{\vect k} (i\varepsilon_l) G_{\vect k}(i\varepsilon_l + i \hbar \Omega_l) 
 G_{\vect k'} (i\varepsilon_{l'}) G_{\vect k'}(i\varepsilon_{l'} + i \hbar \Omega_l) 
\bigl\{ G_{\vect k''}(i\varepsilon_{l''}+i \hbar \Omega_l) - G_{\vect k''}(i\varepsilon_{l''}) \bigr\} \nonumber \\
&\times&\!\!\! \Bigl\{ - D_{\vect k-\vect k''}(i\varepsilon_l - i \varepsilon_{l''}) D_{\vect k'-\vect k''}(i\varepsilon_{l'} - i \varepsilon_{l''}) 
{\cal F}_3 k_x^2 {k'_y}^2 \bigl( {\cal F}_0 + {\cal F}_1 \vect k \cdot \vect k'' \bigr) \bigl( {\cal F}_0 + {\cal F}_1 \vect k' \cdot \vect k'' \bigr) \nonumber \\
&\times&\!\!\! e^{i(\vect k-\vect k'')\cdot (\vect R_{n'} - \vect R_n ) + i(\vect k'-\vect k'') \cdot (\vect R_n - \vect R_{n''})} \nonumber \\
& +&\!\!\! 2 D_{\vect k-\vect k'}(i\varepsilon_l - i \varepsilon_{l'}) D_{\vect k''-\vect k'}(i\varepsilon_{l''} - i \varepsilon_{l'}) 
{\cal F}_3 k_x^2 {k''_y} {k'_y} \bigl( {\cal F}_0 + {\cal F}_1 \vect k \cdot \vect k' \bigr) \bigl( {\cal F}_0 + {\cal F}_1 \vect k' \cdot \vect k'' \bigr) \nonumber \\
&\times&\!\!\! e^{i(\vect k-\vect k')\cdot (\vect R_{n'} - \vect R_n ) + i(\vect k''-\vect k') \cdot (\vect R_{n} - \vect R_{n''})} \Bigr\} .
\label{eq:sigmaSS0}
\end{eqnarray}
Here, the last term in Eq.~(1) is not considered because this term is proportional to $(\vect J_n \cdot \vect J_n) \approx const.$. 
The Matsubara summation can be carried out similarly as in the side-jump mechanism, leading to 
\begin{eqnarray}
\sigma^{skew \, scat.}_{SH} \!\!\!&=&\!\!\! \frac{e^2 \hbar^3}{\pi m^2 V N^5} \! \int \!\! d\varepsilon d \omega d \omega' \!\! \sum_{n,n',n''} \sum_{\vect k, \vect k', \vect k''} 
\nonumber \\
&\times&\!\!\! \biggl[ 
{\cal F}_3 k_x^2 {k'_y}^2 \bigl( {\cal F}_0 + {\cal F}_1 \vect k \cdot \vect k'' \bigr) \bigl( {\cal F}_0 + {\cal F}_1 \vect k' \cdot \vect k'' \bigr) 
B_{\vect k - \vect k''}(\omega) \Bigl\{ b(\omega) + f \bigl(\varepsilon_{\vect k} \bigr) \Bigr\} 
B_{\vect k' - \vect k''}(\omega') \Bigl\{ b(\omega') + f \bigl(\varepsilon_{\vect k'} \bigr) \Bigr\} \nonumber \\
&\times&\!\!\! \biggl\{ \partial_\varepsilon f(\varepsilon) 
\bigl| G_{\vect k}^R(\varepsilon + \omega) \bigr|^2
\bigl| G_{\vect k'}^R(\varepsilon + \omega') \bigr|^2 
\Im \bigl\{ G_{\vect k''}^R(\varepsilon) \bigr\} 
+ f(\varepsilon) \Im \Bigl\{ \bigl(G_{\vect k}^R(\varepsilon + \omega) G_{\vect k'}^R(\varepsilon+\omega') \bigr)^2
\partial_\varepsilon G_{\vect k''}^R(\varepsilon) \Bigr\} \biggr\} \nonumber \\
&\times&\!\!\! e^{i(\vect k-\vect k'')\cdot (\vect R_{n'} - \vect R_n ) + i(\vect k'-\vect k'') \cdot (\vect R_n - \vect R_{n''})} \nonumber \\
&-&\!\!\! 2 
{\cal F}_3 k_x^2 {k''_y} {k'_y} \bigl( {\cal F}_0 + {\cal F}_1 \vect k \cdot \vect k' \bigr) \bigl( {\cal F}_0 + {\cal F}_1 \vect k' \cdot \vect k'' \bigr) 
 B_{\vect k - \vect k'}(\omega) \Bigl\{ b(\omega) + f \bigl(\varepsilon_{\vect k} \bigr) \Bigr\} 
B_{\vect k'' - \vect k'}(\omega') \Bigl\{ b(\omega') + f \bigl(\varepsilon_{\vect k''} \bigr) \Bigr\} \nonumber \\
&\times&\!\!\! \biggl\{ \partial_\varepsilon f(\varepsilon) 
\bigl| G_{\vect k}^R(\varepsilon + \omega) \bigr|^2
\bigl| G_{\vect k'}^R(\varepsilon) \bigr|^2
\Im \bigl\{ G_{\vect k''}^R(\varepsilon+\omega') \bigr\} 
+f(\varepsilon) \Im \Bigl\{ \bigl(G_{\vect k}^R(\varepsilon + \omega) G_{\vect k'}^R(\varepsilon) \bigr)^2
\partial_\varepsilon G_{\vect k''}^R(\varepsilon+\omega') \Bigr\} \biggr\} \nonumber \\
&\times&\!\!\! e^{i(\vect k-\vect k')\cdot (\vect R_{n'} - \vect R_n ) + i(\vect k''-\vect k') \cdot (\vect R_{n} - \vect R_{n''})}
\biggr]. 
\label{eq:sigmaSS1}
\end{eqnarray}

Again, we focus on the Fermi surface terms which are proportional to $\partial_\varepsilon f(\varepsilon)$. 
Carrying out $\varepsilon$, $\omega$ and $\omega'$ integrals, one obtains 
\begin{eqnarray}
\sigma^{skew \, scat.}_{SH} \!\!\!&\approx&\!\!\! \frac{4 e^2 \hbar \pi^2 \tau_c^2}{m^2 V N^5} \!\! \sum_{n,n',n''} \sum_{\vect k, \vect k', \vect k''} 
\delta(\varepsilon_{\vect k''}) 
B_{\vect k - \vect k''}(\varepsilon_{\vect k}) \Bigl\{ b(\varepsilon_{\vect k}) + f \bigl(\varepsilon_{\vect k} \bigr) \Bigr\} 
B_{\vect k' - \vect k''}(\varepsilon_{\vect k'}) \Bigl\{ b(\varepsilon_{\vect k'}) + f \bigl(\varepsilon_{\vect k'} \bigr) \Bigr\} \nonumber \\
&\times&\!\!\! \bigl\{ 
- {\cal F}_3 k_x^2 {k'_y}^2 \bigl({\cal F}_0 + {\cal F}_1 \vect k \cdot \vect k''\bigr) \bigl ({\cal F}_0 + {\cal F}_1 \vect k' \cdot \vect k'' \bigr) 
+2 
{\cal F}_3 k_x^2 {k''_y} {k'_y} \bigl( {\cal F}_0 + {\cal F}_1 \vect k \cdot \vect k'' \bigr) \bigl({\cal F}_0 + {\cal F}_1 \vect k' \cdot \vect k'' \bigr) 
\bigr\} \nonumber \\
&\times&\!\!\! e^{i(\vect k-\vect k'')\cdot (\vect R_{n'} - \vect R_n ) + i(\vect k'-\vect k'') \cdot (\vect R_n - \vect R_{n''})} .
\label{eq:sigmaSS5}
\end{eqnarray}

As in the side-jump case, main contributions are from small $|\vect k -\vect k''|$ and small $|\vect k' -\vect k''|$ regions. 
Thus, expanding $\vect k$ and $\vect k'$ from $\vect k''$ as $\vect k = \vect k'' + \vect q$ and $\vect k' = \vect k''  + \vect q'$ 
approximating $e^{i(\vect k-\vect k'')\cdot (\vect R_{n'} - \vect R_n ) + i(\vect k'-\vect k'') \cdot (\vect R_n - \vect R_{n''})} $ by 1, 
and replacing $\vect q$ integrals by $I(T,\delta)$ defined in Eq.~(\ref{eq:F}), 
one arrives at Eq.~(5). 

\subsection{Detail of $I(T,\delta)$}

Focusing on low temperature regimes where the linear approximation $\varepsilon_{\vect k'}\approx \hbar \vect v_F \cdot \vect q$ is justified, 
we further approximate $ b\bigl( \hbar \vect v_F \cdot \vect q\bigr) + f \bigl(\hbar \vect v_F \cdot \vect q \bigr) \approx T/ \hbar \vect v_F \cdot  \vect q$ to arrive at 
\begin{eqnarray}
I(T,\delta)
\approx \frac{1}{N} \sum_{\vect q}
\frac{T/\Gamma_q}{\bigl( \delta+A q^2\bigr)^2 + \bigl( \hbar \vect v_F \cdot \vect q/\Gamma_q \bigr)^2}.
\end{eqnarray}
Considering a three dimensional system, the $\vect q$ integral is evaluated as 
\begin{eqnarray}
I(T,\delta)
&\approx&\!\!\!
\frac{a^3}{(2\pi)^3} \int_0^{T/\hbar v_F} \!\!\!\! dq \int_{0}^\pi  \!\!\! d\theta 
\frac{ 2 \pi q^2 T \sin \theta /\Gamma_q}{\bigl( \delta+A q^2\bigr)^2 + \bigl(\hbar v_F q \cos \theta /\Gamma_q \bigr)^2} \nonumber \\
&=&\!\!\! 
\frac{a^3}{(2\pi)^2}\int_0^{T/\hbar v_F} \!\!\!\! dq \frac{T}{\hbar v_F} \frac{2 q}{\delta + A q^2}
\arctan \frac{\hbar v_F q/\Gamma_q}{\delta + A q^2} ,
\label{eq:qint}
\end{eqnarray}
with $a$ being the lattice constant. 

Now, we consider limiting cases, where the analytic form of $I(T,\delta)$ is available. 

(1) {\it Clean metals 
at low temperatures where $\delta + A (T/\hbar v_F)^2 \ll \hbar v_F/\Gamma$}. 
In this case, $\arctan$ is approximated as $\pi/2$, 
leading to 
\begin{eqnarray}
I(T,\delta)
\approx 
\frac{a^3}{(2\pi)^2}\int_0^{T/\hbar v_F} \!\!\!\! dq \frac{T}{A \hbar v_F} \frac{\pi q}{\delta + A q^2} 
=
\frac{1}{8\pi} \frac{a^3 T}{A \hbar v_F} \ln \Biggl[ 1+\frac{A}{\delta} \biggl( \frac{T}{\hbar v_F} \biggr)^2 \Biggr] 
\approx  \frac{1}{8\pi \delta} \biggl( \frac{a T}{\hbar v_F}\biggr)^3. 
\label{eq:qint2}
\end{eqnarray}
Near the FM critical point, $\delta$ is scaled as  $\delta \propto T^{4/3}$ \cite{SMoriya1985}. 
Therefore, $I(T,\delta)$ is expected to be scaled as $T^{5/3}$

(2) {\it  Clean metals at high temperatures with $\delta + A (T/\hbar v_F)^2 \gg \hbar v_F/\Gamma$}.  
In this case, we expand the argument of $\arctan$ to get 
\begin{eqnarray}
I(T,\delta)
\approx 
\frac{a^3}{(2\pi)^2}\int_0^{T/\hbar v_F} \!\!\!\! dq \frac{T}{\hbar v_F} \frac{2 q \hbar v_F /\Gamma}{\bigl( \delta + A q^2 \bigr)^2} 
=
\frac{1}{(2\pi)^2}\frac{\hbar v_F}{\Gamma} \biggl( \frac{a T}{\hbar v_F}\biggr)^3 \frac{1}{\delta \bigl\{ \delta + A (T/\hbar v_F)^2 \bigr\}}. 
\label{eq:qint3}
\end{eqnarray}
At such high temperatures, $\delta$ is linearly proportional to $T$ \cite{SMoriya1985,SUeda1975}. 
Therefore, $I(T,\delta)$ is expected to be proportional to $T$. 

(3) {\it  Dirty metals at low temperatures with $T  \ll \Gamma q_c \delta$}.  
In this case, we take $\Gamma_{q}=\Gamma q_c$ and expand the argument of $\arctan$ to get 
\begin{eqnarray}
I(T,\delta)
\!\!\!&\approx &\!\!\!
\frac{a^3}{(2\pi)^2}\int_0^{T/\hbar v_F} \!\!\!\! dq \frac{T}{\hbar v_F} \frac{2 q^2 \hbar v_F /\Gamma q_c}{\bigl( \delta + A q^2 \bigr)^2} 
=
\frac{a^3}{(2\pi)^2} \frac{\hbar v_F}{\Gamma q_c} \frac{T}{\hbar v_F} 
\Biggl[ \frac{1}{A \sqrt{A \delta}}\arctan \sqrt{\frac{A}{\delta}}q -\frac{q}{A \bigl(\delta + Aq^2 \bigr)}
\Biggr]_0^{T/\hbar v_F}
\nonumber \\
\!\!\!&\approx &\!\!\!\frac{1}{(2\pi)^2}\frac{\hbar v_F}{a \Gamma q_c} \biggl( \frac{a T}{\hbar v_F}\biggr)^4 \frac{1}{\delta \bigl\{ \delta + A (T/\hbar v_F)^2 \bigr\}}. 
\label{eq:qint4}
\end{eqnarray}
Since the damping $\Gamma_q$ is independent of $q$ in this temperature regime, 
$\delta$ is scaled as $\delta \propto T^{3/2}$ \cite{SNagaosa1999}. 
Therefore, $I(T,\delta)$ is expected to be proportional to $T$.

(4) {\it  Dirty metals at moderately high temperatures with $T  >  \Gamma q_c \delta$}.  
In this case, we expand the argument of $\arctan$ and separate the $q$ integral into two regions, $q \leq q_c$ and $q \geq q_c$. 
This leads to  
\begin{eqnarray}
I(T,\delta)
\!\!\!&\approx &\!\!\!
\frac{a^3}{(2\pi)^2} \frac{T}{\hbar v_F} \left\{ 
\int_0^{q_c} \!\!\!\! dq  \frac{2 q^2 \hbar v_F /\Gamma q_c}{\bigl( \delta + A q^2 \bigr)^2} 
+
\int_{q_c}^{T/\hbar v_F} \!\!\!\! dq \frac{2 q \hbar v_F /\Gamma}{\bigl( \delta + A q^2 \bigr)^2}
\right\} \nonumber \\
\!\!\!&=&\!\!\!
\frac{a^3}{(2\pi)^2} \frac{T}{\hbar v_F} \left\{ 
\frac{\hbar v_F}{\Gamma q_c} \Biggl[ \frac{1}{A\sqrt{A \delta}}\arctan \sqrt{\frac{A}{\delta}}q -\frac{q}{A \bigl(\delta + Aq^2 \bigr)}
\Biggr]_0^{q_c}
-
\frac{\hbar v_F}{\Gamma} \Biggl[ \frac{1}{A\bigl\{ \delta + A q^2 \bigr\}}\Biggr]_{q_c}^{T/\hbar v_F} \right\}\nonumber \\
\!\!\!&=&\!\!\!
\frac{a^3}{(2\pi)^2} \frac{T}{\hbar v_F} \left\{ 
\frac{\hbar v_F}{\Gamma q_c} \Biggl[ \frac{1}{A\sqrt{A \delta}}\arctan \sqrt{\frac{A}{\delta}}q_c -\frac{q_c}{A \bigl(\delta + Aq_c^2 \bigr)}
\Biggr]
-
\frac{\hbar v_F}{\Gamma} \Biggl[ \frac{1}{A\bigl\{ \delta + A (T/\hbar v_F)^2 \bigr\}} -\frac{1}{A\bigl\{ \delta + A q_c^2 \bigr\}} 
\Biggr] \right\}
\nonumber \\
\!\!\!&\approx&\!\!\!
\frac{1}{(2\pi)^2}
\frac{\hbar v_F}{\Gamma} \biggl( \frac{a T}{\hbar v_F}\biggr)^3  \frac{1}{\delta \bigl\{ \delta + A (T/\hbar v_F)^2 \bigr\}} 
\label{eq:qint5}
\end{eqnarray}
The final form is the same as Eq.~(\ref{eq:qint3}). 
In this temperature regime, $\delta$ is proportional to $T$ \cite{SMoriya1985,SUeda1975}. 
Therefore $I(T,\delta)$ is also proportional to $T$. 

\subsection{Carrier lifetime by the spin fluctuation} 

\begin{figure}
\begin{center}
\includegraphics[width=0.2\columnwidth, clip]{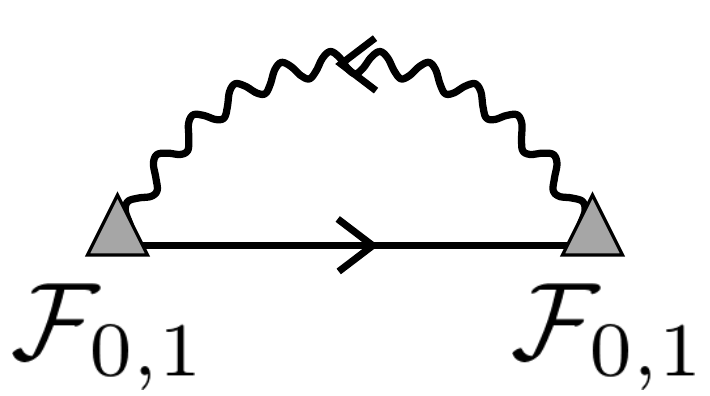}
\caption{Diagrammatic representation for the electron self-energy. 
}
\label{fig:selfenergy}
\end{center}
\end{figure}

Here, we consider the electron self-energy $\Sigma_{\vect k} (i \varepsilon_l)$ due to the coupling with the spin fluctuation. 
The lowest order self-energy is given by (see Fig.~\ref{fig:selfenergy} for the diagramatic representation)
\begin{eqnarray}
\Sigma_{\vect k}(i \varepsilon_l) = \frac{T}{N^3} \sum_{n,n'} \sum_{\vect q, i\omega_{l'}} e^{i (\vect k + \vect q) \cdot (\vect R_n -\vect R_{n'})}
\Bigl\{ {\cal F}_0 + 2 {\cal F}_1 \bigl( k^2 + \vect k \cdot \vect q \bigr) \Bigr\}^2
G_{\vect k + \vect q}(i \varepsilon_l + i \omega_{l'}) D_{\vect q} (i \omega_{l'}). 
\end{eqnarray} 
After carrying out the Matsubara summation and the analytic continuation $i \varepsilon_l \rightarrow \varepsilon + i \eta$ 
with $i\eta$ being a small imaginary number, 
the imaginary part of the self-energy becomes 
\begin{eqnarray}
\Im \Sigma_{\vect k} (\varepsilon) \approx - \frac{\pi}{N^3} \sum_{\vect q} \int d \omega 
e^{i (\vect k + \vect q) \cdot (\vect R_n -\vect R_{n'})}
\Bigl\{ {\cal F}_0 + 2 {\cal F}_1 \bigl( k^2 + \vect k \cdot \vect q \bigr) \Bigr\}^2
B_{\vect q} (\omega)
\bigl\{b(\omega) + f(\omega + \varepsilon) \bigr\}
\delta(\varepsilon+\omega-\varepsilon_{\vect k + \vect q}). 
\end{eqnarray}

As in the SH conductivity, 
we focus on the low-energy part $\varepsilon=0$, approximate $\varepsilon_{\vect k + \vect q} \approx \hbar {\vect v}_F \cdot \vect q$ and 
$e^{i (\vect k + \vect q) \cdot (\vect R_n -\vect R_{n'})} \approx 1$. 
This leads to
\begin{eqnarray}
\Im \Sigma_{\vect k} (0) \approx - \frac{\pi}{N} n_m^2 \sum_{\vect q} \Bigl\{ F_0 + 2 F_1 \bigl( k^2 + \vect k \cdot \vect q \bigr) \Bigr\}^2
B_{\vect q} (\hbar {\vect v}_F \cdot \vect q)
\bigl\{b(\hbar {\vect v}_F \cdot \vect q) + f(\hbar {\vect v}_F \cdot \vect q) \bigr\}. 
\end{eqnarray}

Neglecting the small contribution from $\vect k \cdot \vect q$, one obtains for $k=k_F$
\begin{eqnarray}
\Im \Sigma_{k_F} (0)  \approx -n_m^2 \bigl( F_0 + 2 F_1  k_F^2 \bigr)^2 I (T,\delta) = - \hbar / 2 \tau_{sf}. 
\end{eqnarray}

\subsection{Maximum $\sigma_{SH}^{side \, jump}$ and $\sigma_{SH}^{skew \, scat.}$} 

Here, we consider $\tau_c I(T,\delta)$ at low temperatures in the clean limit. 
Parameterizing the carrier lifetime by the spin fluctuation as $\tau_{sf}^{-1}=\tau_{sf,0}^{-1} (T/T_{sf})^{5/3}$ and $I(T,\delta)=\alpha (T/T_{sf})^{5/3}$, 
where $T_{sf}=\hbar v_F/a$, we differentiate $\tau_c I(T,\delta)$ with $\tau_c^{-1}=\tau_{sf}^{-1}+\tau_{dis}^{-1}+\tau_{ee}^{-1}+ \tau_{ep}^{-1}$ with respect to $T$ 
as 
\begin{eqnarray}
\frac{d \tau_c I(T,\delta)}{dT} \!\!\!&=&\!\!\! \alpha \tau_c^2 \Biggl[ \frac{5}{3 T_{sf}} \biggl(\frac{T}{T_{sf}}\biggr)^{2/3} \tau_c^{-1} 
- \biggl(\frac{T}{T_{sf}}\biggr)^{5/3} \Biggl\{ 
\tau_{sf,0}^{-1} \frac{5}{3 T_{sf}} \biggl(\frac{T}{T_{sf}}\biggr)^{2/3}
+ \tau_{ee,0}^{-1} \frac{2}{T_F} \biggl(\frac{T}{T_{F}}\biggr)
+ \tau_{ep,0}^{-1} \frac{5}{T_D} \biggl(\frac{T}{T_{D}}\biggr)^{4}
\Biggr\}
\Biggr] \nonumber \\
\!\!\!&=&\!\!\!
\frac{5 \alpha \tau_c^2}{3 T_{sf}}\biggl(\frac{T}{T_{sf}}\biggr)^{2/3} \Biggl[
\tau_{dis}^{-1} - \frac{1}{5} \tau_{ee,0}^{-1} \biggl(\frac{T}{T_{F}}\biggr)^2 - 2 \tau_{ep,0}^{-1} \biggl(\frac{T}{T_{D}}\biggr)^5
\Biggr]=0. 
\end{eqnarray}
An approximate solution for this equation is
$T_{max}=T_F (5 \tau_{ee,0}/\tau_{dis})^{1/2}$ for $T_F \ll T_D$ or 
$T_{max}=T_D (\tau_{ep,0}/2 \tau_{dis})^{1/5}$ for $T_F \gg T_D$. 

At this temperature, $\tau_c I(T,\delta)$ becomes 
\begin{eqnarray}
\tau_c I(T,\delta) = \alpha \biggl( \frac{T_F}{T_{sf}}\biggr)^{5/3} \biggl( \frac{5\tau_{ee,0}}{\tau_{dis}}\biggr)^{5/6}
\Biggl\{6 \tau_{dis}^{-1} + \tau_{sf,0}^{-1}  \biggl( \frac{T_F}{T_{sf}}\biggr)^{5/3} \biggl( \frac{5\tau_{ee,0}}{\tau_{dis}}\biggr)^{5/6}
\Biggr\}^{-1}
\end{eqnarray}
for $T_F \ll T_D$, or
\begin{eqnarray}
\tau_c I(T,\delta) = \alpha \biggl( \frac{T_D}{T_{sf}}\biggr)^{5/3} \biggl( \frac{\tau_{ep,0}}{2 \tau_{dis}}\biggr)^{1/3}
\Biggl\{\frac{3}{2} \tau_{dis}^{-1} + \tau_{sf,0}^{-1}  \biggl( \frac{T_D}{T_{sf}}\biggr)^{5/3} \biggl( \frac{5\tau_{ep,0}}{\tau_{dis}}\biggr)^{1/3}
\Biggr\}^{-1}
\end{eqnarray}
for $T_F \gg T_D$. 

When $\tau_{dis}^{-1} \ll \tau_{sf,0}^{-1}$, $\tau_c I(T,\delta) \approx \alpha \tau_{sf,0}$. 
This leads to Eq.~(7). 

Similar analysis can be done for the dirty limit. 
The expression for $T_{max}$ is the same as the clean limit. 
However, the leading term of $\sigma_{SH,max}^{side \, jump}$ explicitly depends on both $\tau_{dis}$ and $\tau_{ee,0}$ or $\tau_{ep,0}$. 

Skew scattering contribution $\sigma_{SH}^{skew \, scat.}$ is proportional to $\tau_c^2 F^2(T,\delta)$. 
Therefore, $\sigma_{SH}^{skew \, scat.}$ is expected to be maximized  to be $\sigma_{SH, max}^{skew \, scat.}$ in Eq.~(8) 
at the same temperature $T_{max}$ as $\sigma_{SH}^{side \, jump}$ .

\end{document}